\documentclass[conference]{IEEEtran}
\usepackage[dvips]{graphicx}
\usepackage{caption}
\usepackage{subcaption}
\usepackage{listings}
\usepackage{mathptmx}

\begin{document}
\title{MOMCC: Market-Oriented Architecture for Mobile~Cloud~Computing Based on Service~Oriented~Architecture}

\author{\IEEEauthorblockN{Saeid Abolfazli$^1$, Zohreh Sanaei$^2$, Abdullah Gani$^3$, Muhammad Shiraz$^4$}
\IEEEauthorblockA{Mobile Cloud Computing Research Lab\\
Faculty of Computer Science and Information Technology\\
University of Malaya, Kuala Lumpur, Malaysia \\
Email: abolfazli$^1$, sanaei$^2$,abdullahgani$^3$@ieee.org, muh\_shiraz@siswa.um.edu.my$^4$ }
}
\maketitle
\begin{abstract}
The vision of augmenting computing capabilities of mobile devices, especially smartphones with least cost is likely transforming to reality leveraging cloud computing. Cloud exploitation by mobile devices breeds a new research domain called Mobile Cloud Computing (MCC). However, issues like portability and interoperability should be addressed for mobile augmentation which is a non-trivial task using component-based approaches. Service Oriented Architecture (SOA) is a promising design philosophy embraced by mobile computing and cloud computing communities to stimulate portable, complex application using prefabricated building blocks called Services. Utilizing distant cloud resources to host and run Services is hampered by long WAN latency. Exploiting mobile devices in vicinity alleviates long WAN latency, while creates new set of issues like Service publishing and discovery as well as client-server security, reliability, and Service availability. In this paper, we propose a market-oriented architecture based on SOA to stimulate publishing, discovering, and hosting Services on nearby mobiles, which reduces long WAN latency and creates a business opportunity that encourages mobile owners to embrace Service hosting. Group of mobile phones simulate a nearby cloud computing platform. We create new role of \textit{Service host} by enabling unskilled mobile owners/users to host Services developed by skilled developers. Evidently, Service availability, reliability, and Service-oriented mobile application portability will increase towards green ubiquitous computing in our mobile cloud infrastructure.
\end{abstract}
\IEEEpeerreviewmaketitle

\section{Introduction}
\IEEEPARstart Vision of performing computing-intensive tasks on the go has been around since long and users are increasingly demanding rich interaction experience similar to (or even better than) stationary computers. However, their resource poverty beside their compact and light nature hinders users' vision. Researchers in academia and industry endeavour to augment mobile devices in different ways which was studied in our prior work \cite{SaeidAbolfazli2012}. We illustrated that mobile augmentation approaches mainly require external resources such as nearby computing devices \cite{Satyanarayanan2001,Satyanarayanan2009} or distant clouds servers\cite{Zhang2011a}. Although nearby computing devices likely offer lower latency, they are weak devices that might not be able to perform complex resource-intensive tasks. Moreover, their services are voluntary and free that gives them freedom to terminate their services anytime. These free computing machines can be an attacking point utilized by an attacker in the absence of monitoring authority. Hence, their usability is obstructed. Cloud infrastructure are offering reliable pay-as-you-use services based on a service-level agreement between service provider and consumer \cite{buyya2009cloud}. So, they are likely a better alternative to those free vulnerable services.

Exploitation of cloud computing by mobile handhelds breeds a new research domain called Mobile Cloud Computing (MCC), which is the state-of-the-art computing paradigm comprised of three heterogeneous domains of mobile computing, cloud computing, and networking \cite{ZohrehSanaei2012}. Such non-uniformity stems several challenges such as portability, interoperability, and fragmentation that are deemed to be alleviated leveraging SOA \cite{Erl2005} as a promising design philosophy embraced by mobile computing and cloud computing communities to stimulate portable, complex application using prefabricated building blocks called Services. Services are prefabricated codes that are developed in languages like Java, .Net, and PHP and are often publicized in a publicly accessible repository to be discovered and invoked by clients to provide specific functionality. Currently, services are developed by skilled developer(s) and hosted on a publicly available server on the Web and recently on the Clouds. However, storing Services on the Web and Cloud infrastructure not only increases long WAN latency and decreases security due to the vulnerability of the channel of Internet, but also squanders noticeable amount of energy from the energy-constraint mobile devices. 

To alleviate challenges of hosting and running services on distant servers, nearby mobile devices like smartphones and Tablets are likely an appropriate alternative if fundamental requirements like security and reliability achieved. More than 86\% of the world population are mobile subscriber and mobile phone market share is rapidly increasing \cite{Gartner2012}. Although their computing ability especially battery is very limited, accumulative power of swarm of mobile devices can turn them into a giant resource-rich, ubiquitous infrastructure to not only provide a low-cost, green distributed computing, but also generate an income source for their owners. Despite of mobile phones' ownership and maintenance cost including subscription, traffic, and energy fee mobile devices are hardly an income source for their owners. The paid Service hosting and execution is deemed to be embraced by mobile owners if issues such as Service publishing and discovery as well as client-server security, reliability, and Service availability are addressed. Services can be hosted on a hosting toolkit like \cite{Asif2007} which is a lightweight hosting toolkit for resource-poor mobile devices. 

In this paper, we propose an approach to publish, discover, and host Services on nearby mobiles, which not only alleviates aforementioned issues, but also creates a business opportunity that encourages mobile owners to embrace Service hosting and execution. We create new role of \textit{Service host} to enable unskilled mobile client to host Services developed by skilled developers. A proper billing system can divide the income according to a negotiated agreement between engaging parties. To the best of our knowledge, this is the first proposal that aims to utilize nearby mobile devices under supervision of a supervisory body to host and run services based on an agreement.

The remainder of this article is organized as follows: we review related works in Section \ref{relatedworks}. Section \ref{SOA} briefly describes Service Oriented Architecture (SOA) followed by Section \ref{MCCarchitecure} that presents the proposed architecture. We explain advantages and disadvantages of our architecture in Section \ref{opportunities} and paper is concluded in Section \ref{conclusions}. 

\section{Related Works} \label{relatedworks}
The concept of utilizing remote resources to augment computing capabilities of mobile devices was firstly introduced by Satyanarayana \cite{Satyanarayanan2001} in pervasive computing. The author visions to host and run the resource-intensive components of the mobile application on a nearby, powerful, stationary computer called surrogates. However, surrogates provide free services and can terminate their services anytime during runtime. They also can violate security and privacy of mobile users in the absence of supervising party. Migrating overhead in the mobile side and virtualization delay in the surrogate side, hinder success of cyber foraging \cite{Sharifi2011}. Moreover for every execution, the code should be offloaded to surrogate machine that increases the communication overhead and network latency. The author and his colleagues later envisioned to utilize nearby computers including mobile devices \cite{Satyanarayanan2009} to overcome the long WAN latency while running the resource-intensive mobile applications. However, security of surrogates and execution latency demand further efforts. 

In another effort \cite{Asif2007} authors develop a lightweight Service hosting toolkit for resource-poor mobile devices with ability to migrate resource-intensive part(s) of the Services to a remote resource-rich computing device. However, identifying resource-intensive part(s) of a Service, allocating remote resources, code offloading, and result collection prolong execution time and dissipate large amount of local resources. To avoid such Service migration overhead, in our Market-Oriented Mobile Cloud Computing (MOMCC) architecture, we restrict mobile hosts to only host those Services that can be executed without offloading. This step reduces the complexity of Service hosting on mobile devices. 

Hyrax \cite{hyrax}, is a Hadoop-based cloud platform consists of several Android smartphones to simulate a nearby cloud of smartphones. Hyrax enables direct communication between mobiles to avoid global network bottleneck and deploys MapReduce approach to breakdown and assign tasks to each smartphone. However, due to software engineering approaches used in Hyrax, there is a tight dependency between code and underlying platform that exacerbates portability and interoperability problems. In \cite{Huerta-Canepa2010}, authors deploy mobile devices to create a virtual cloud computing to facilitate resource-intensive tasks on mobile devices. However, context gathering, resource sensing, and offloading overhead, largely impose overhead on naive mobile devices.

\section{Service Oriented Architecture (SOA)} \label{SOA}
SOA is a design philosophy that follows ultimate aim of reducing development time, cost, and complexity using prefabricated building blocks called Services, while facilitates application maintenance \cite{Papazoglou2007}. In this design philosophy, several sequential or parallel Services are bind together to build a new complex functionality. SOA has been utilized in computing domains such as grid computing \cite{Srinivasan2005} or was cornerstone for technologies such as cloud computing to alleviate several fundamental challenges such as portability, interoperability, and integration of applications and software systems. This is due to the fact that Services are autonomous and platform-neutral meaning that different Services running on heterogeneous platforms can still collaborate toward fulfilling a complex task. Service are providing higher-level abstraction compared with components which makes them suitable technology for large scale computing domains such as MCC. 

To alleviate several problems like portability and interoperability in MCC that are exacerbated by heterogeneity \cite{ZohrehSanaei2012}, SOA is one of the best approaches by which online, platform-neutral, interoperable mobile applications can be built and ported to several mobile platforms with minimal modification and editing. A natural approach to leverage SOA in MCC is to deploy computing Services in the cloud and invoke them in runtime. However, this is impeded by several challenges such as long WAN latency, security risks of utilizing the wireless network as well as the channel of Internet, and energy deficiency of mobile clients.

An alternative approach to utilize SOA in MCC is to deploy Services on nearby mobile devices to be invoked with reduced latency (utilizing alternative communication technologies such as cellular networks and Wi-Fi can enhance performance) without accessing the risky channel of Internet. Nevertheless, there is no publicly acceptable incentive and motivation neither for Service developer nor mobile hosts (except volunteers who are willing to collaborate freely). Developers are required to build publicly available, reusable Services to be utilized by other programmers in creating complex systems and applications, while mobile owners are needed to lease their computational resources such as CPU, memory, and most importantly battery to host and run the Services. In order to encourage the public and realize the vision of utilizing resources of nearby mobile devices (such as smartphones and Tablets), we propose a market-oriented architecture in which Service developers, brokers, and hosts can earn for their public Services whereas Service consumers should pay as they use (similar to the concept of Cloud computing and utility).

\section{Market-Oriented Mobile Cloud Architecture} \label{MCCarchitecure}
\begin{figure*}
\begin{subfigure}[b]{0.55\textwidth}
\centering
\includegraphics[width=\textwidth]{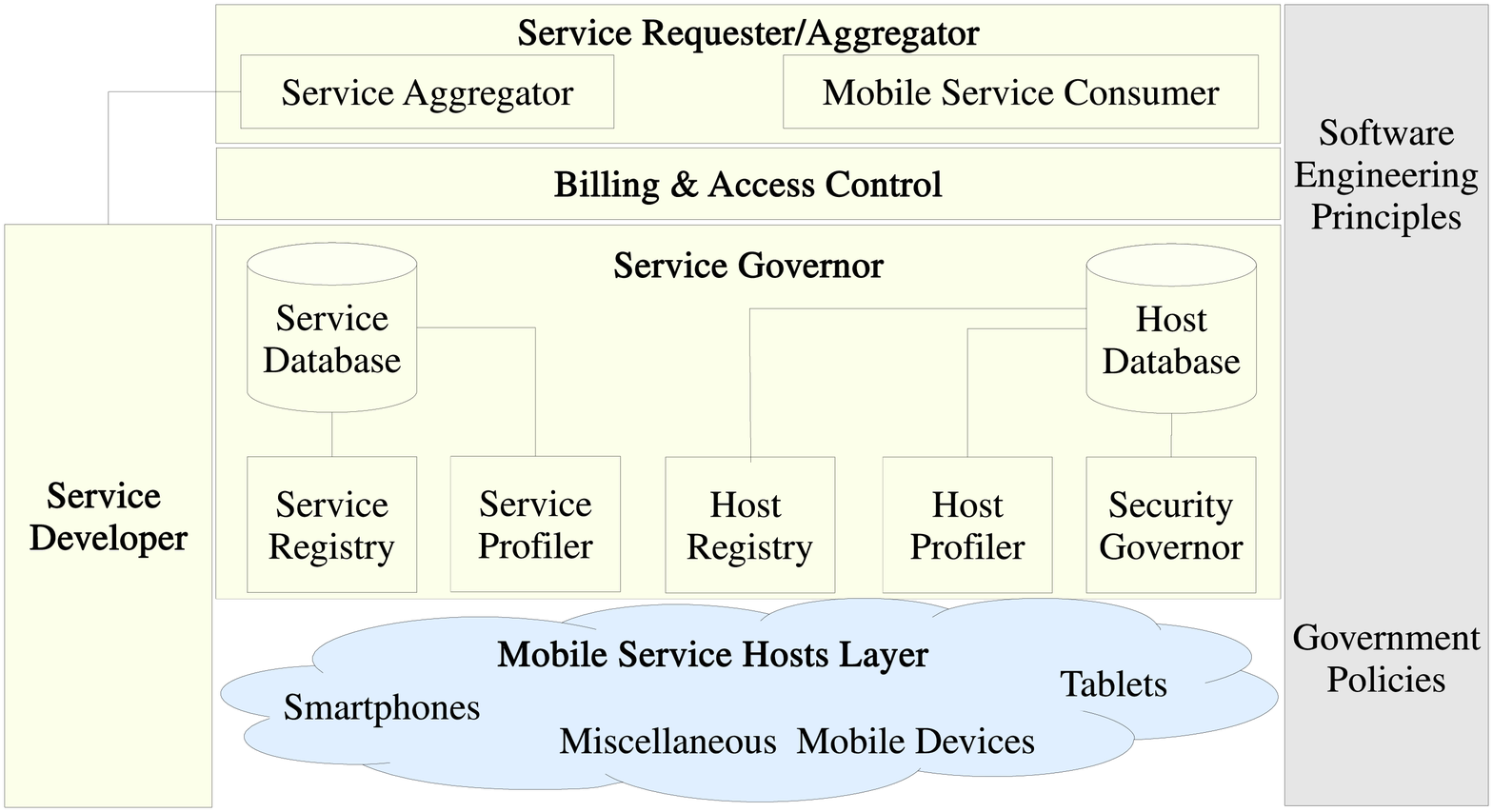}
\caption{Layered Architecture of MOMCC}
\label{layers}
\end{subfigure}%
\qquad
\begin{subfigure}[b]{0.38\textwidth}
\centering
\includegraphics[width=\textwidth]{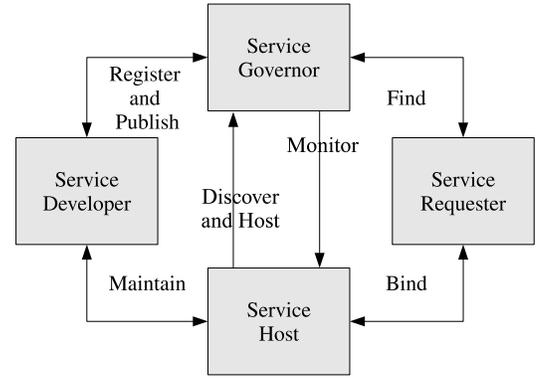}
\caption{The Block Diagram of MOMCC}
\label{roles}
\end{subfigure}
\caption{The Architecture of Market-Oriented Mobile Cloud Computing (MOMCC)}\label{architecture}
\end{figure*}

Figure \ref{layers} depicts our proposed layered architectural model consists of four entities, namely Service developer, governor, host, and requester/aggregator. The functional relationship between these four building blocks are depicted in Figure \ref{roles}. Initially, the Service developer registers and publishes its Service(s) to the Service governor which plays the role of UDDI for Web Services. Service host communicates with the Service governor to browse available Services and request for hosting. In runtime, the requester or aggregator (program developer who aggregates services to quickly build new composite applications) will query the required Service against the Service governor to identify the nearby Service host. Once found, the requester can directly invoke and bind the Service. Service developer has also direct link with the Service host to maintain and update the Service if required. Service governor monitors the performance and reliability of the Service host for future desicions. In this architecture there is no direct communication between service requester and developer. Anonymity of Service requester will likely discourage attackers and likely protect privacy of hosts against potential malicious Service developer. 

\subsection{Service Developer}
Service developer is an organization or individual developer responsible for design and development of the specific Service offered to the requesters. Service developers are able to earn money when their publicly defined Services are available and utilized by consumer. Service complexity varies from a small mathematical function to a complex enterprise task. However, Services should be lightweight building blocks executable on resource-constrained mobile devices with least possible footprint. In MOMCC, resource-intensive Services will be broken down into small sub-Services to avoid runtime code migration and offloading. However, invoking overhead is an important factor while defining Service granularity, because invoking fine grained Services imposes excessive processing and communication overheads that prolongs overall execution time leads to Service consumer's resource drainage. 

\subsection{Billing and Access Control}
In this architecture, Service developing and hosting are paid Services. Hence, billing and access control block is responsible to control and audit service consumption, maintain billing process, and negotiate among service developers, hosts, and requesters to establish a mutual agreement similar to the cloud Service-level agreement \cite{buyya2009cloud}. The negotiations happen in various stages. Initially, Service provider can negotiate with the billing unit at the time of registration, while Service host can negotiate prior to Service hosting. In this case, Service hosts neither negotiate with developer nor requester and consume native resources for Service execution only.

\subsection{Service Governor}
Service governor is an entity located on a centralized server responsible for monitoring and supervision tasks. Considering huge number of Services, developers, hosts, and requesters the need for a supervising and monitoring entity is vital for the success of the whole system. Service governor is the main governing entity in MOMCC with several crucial responsibilities that are briefed as follows:

\noindent \textit{- Service Registry:} Service registry acts as a public Service repository similar to UDDI (Universal Description Discovery Integration). Service registry maintains a local database called 'Service Database' to store and retrieve available Service descriptions, corresponding providers, and hosting entities. In order to enhance security of both mobile host and service consumer, the Service code is scrutinized against malicious codes by the service registry. When a programmer develops a Service, the Service should be registered with Service registry to be exposed to future programmers. While registering a Service as a business entity, Service developers state Service functionality, input, output, binding method, security level, and minimum required hardware or platform by exchanging a SOAP (simple Object Access Protocol) message with the Service registry. A part of a sample SOAP message to declare minimum requirements is shown in Listing 1. From security point of view, Services can be classified into various categories like low, medium, and high depends on the nature of their functionality and engaging data. Service registry also is responsible to check the service database to find the requested Service (using Service name or its description) and reply to the requesters.

\begin{small}
\lstset{language=XML}

\begin{lstlisting} [caption={Service requirement specification in a sample SOAP message}]

<?xml version="1.0" encoding="UTF-8" ?>
...
<HostRequirments>
  <Platform>
    <OS>Android</OS>
    <MinVersion>3.2</MinVersion>
  </Platform>
  <MinRequiredResources>
    <CPU>512</CPU>
    <Memory>2</Memory>
    <Storage>5</Storage>
    <Energy>500</Energy>
  </MinRequiredResources>
</HostRequirments>
\end{lstlisting}
\end{small}

\noindent \textit{- Service Profiler:} Service profiler monitors the overall functionality and performance of various Services. Service provider is responsible to maintain the Service in case of any malfunction. However, to enhance the quality of the Services, their overall functionality, availability, and vulnerability is monitored by Service profiler to substitute low quality services with more efficient ones if required. 

\noindent \textit{- Host Registry:} Every host should communicate with the public Service registry to browse available Services to be hosted. The host must choose the Services with less resource requirement than its available resources. For example, a service with 2 MB memory need cannot be hosted on a device with 1 MB memory. The host registry is responsible to validate host demands and refuse inappropriate allocation requests. Once the host allocation request is validated, the Service registry will communicate with the Security governor to evaluate host's Security Certificate (SC). The equal or higher SC can be considered acceptable which means, a medium-sensitivity Service can be executed either on a medium or high secure host. For every host, all hosting Services are recorded in the host database and will be utilized to address Service requests.

\noindent \textit{- Host Profiler:} Upon registration of a mobile device as host, the code will be accessible to the device to be hosted locally. For every Service request, the binding procedure is undertaken and Service can be invoked by the mobile Service consumer. The history of Services hosted in every mobile host, including its overall performance, availability, reliability, and security will be collected (using a received execution report after each attempt) and stored in a host database to be utilized for periodic efficiency assessment of each host. The host database is the shared data storage between host registry, profiler, and security governor.

\noindent \textit{- Security Governor} Security is one of the most important concerns among Service consumers, especially in the wireless domain. In this architecture, mobile hosts receives a SC upon successful registry. The trust between security governor and mobile host can be achieved in various ways like reputation trust or identity trust. According to the reputation trust \cite{Josang2007}, a nascent mobile host will be issued the lowest security trust to host low-sensitive Services only due to lack of reputation and prior experience. For every Service execution, the overall behaviour of the host will be monitored and captured by the host profiler, which will be used to promote or demote host's SC. The Service consumer also can rate the quality of Service, which would be an incentive for good hosting. The better and longer execution history, the higher degree of trust. However, this model is subject to sudden change in device/owner's behaviour and might lead to security violation in the absence of solid identity of the host. An alternative trust model can be based on the identity of device holder in the presence of an authentication system. At the time of registering for a Service, the device owner's identity will be verified to discourage possible security attack and a set of credential will be issued. For cellular clients such as smartphones, the SIM card detail can be exploited as an identity and authentication token \cite{rust2008sim}.

\subsection{Mobile Service Host}
A mobile host is a mobile device like smartphone or Tablet which is able and desired to host the implemented Service code and execute it for Service requester on demand. The host activity is a trade-off between resource and money. In order to host a Service, a mobile device must communicate with the Service Governor to find the appropriate Services according to its computing abilities and benefits. In this architecture, Service hosting is a paid Service, hence, Service governor should negotiate with developer (if wishes to provide paid Services), requester, and host to agree on a certain revenue percentage. In generic SOA, Service provider implements or purchase Service implementation, supply its description, and provide technical and business maintenance \cite{Papazoglou2003}. However, due to lack of mobile owners/users' technical competency, we argue that majority of technical and business maintenance of the Services should be accomplished by the Service provider. Every functional and non-functional failure should be reported to the Service provider for escalation. 

Beside finacial benefits of hosting a Service, mobile device will become voulnerable to attack and privacy violation. To avoid such threats, certain mechanisms such as  Sandboxing and Service signing (signing Service code by the service governor) can be deployed. 

\subsection{Service Requester/Aggregator}
\begin{figure}[!ht]
\centering
\includegraphics[scale=0.40]{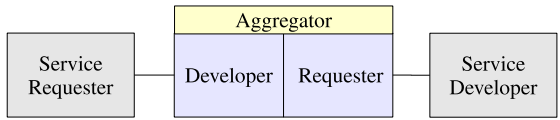} 
\caption{The role of Service Aggregator} \label{aggregator}
\end{figure}

The ultimate aim of SOA is to reduce development time, cost, and complexity using prefabricated building blocked called Services, while facilitates application maintenance. In this design philosophy, several sequential or parallel prefabricated Services are bound together to provide a new complex functionality. Service requester can be the mobile end-consumers of paid Services (there might be some free Services as well) that augments its device processing abilities, Service aggregator. Service aggregator is a Service requester that hosts another Service. Figure \ref{aggregator} depicts relationship between Service requester, aggregator, and provider.

\noindent \textit{Software Engineering Enforcement:}The service governor might enforce software engineering principals to increase reusability, portability, and interoperability while assuring resource efficiency of Services for mobile host. Due to resource limitation of mobile devices, Services should be developed with minimum footprint to consume fewer resources from the hosting devices. Excessive resource consumption by a weakly-designed Service causes extra cost and ultimately will be removed from the popular Service stack.

\noindent \textit{Law Enforcement:} In near future, governments' rules and regulation such as tax and insurance policies as well as QoS metrics are expected to be enforced on Service developers and hosts, which can be performed by Service governor. 

\begin{figure*}[!ht]
\begin{center}
\includegraphics[width=5in, height=2.33in]{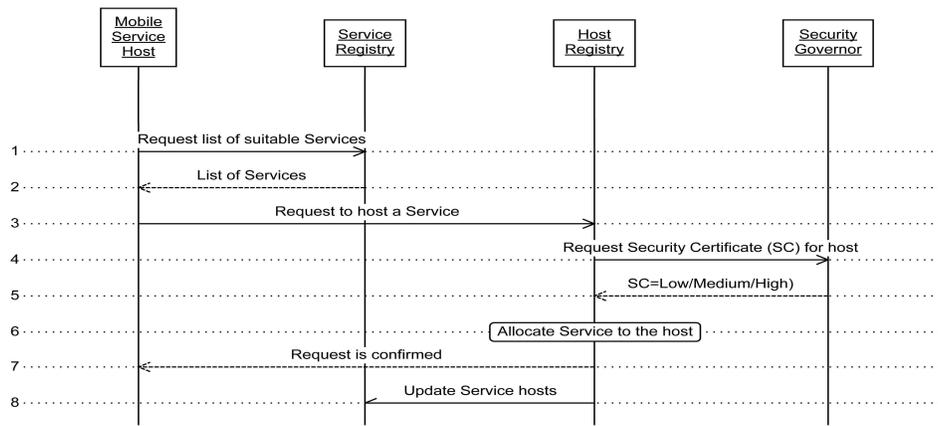} 
\caption{The Collaborative Scenario among Major Entities} \label{sequence}
\end{center}
\end{figure*}

A scenario is illustrated in Figure \ref{sequence}. A 'host' requests a list of available Services to decide the best available option(s) in terms of resource consumption and financial benefits. A list of available Services will be sent back to the 'host' as a reply. Upon deciding a particular Service, 'host' sends a message to the 'host registry' to request hosting of the Service. In order to confirm the 'host' request, the 'host registry' component should verify security level of Service and the SC of the 'host'. Hence, the 'host registry' sends a request to the 'security governor' for the SC of the corresponding 'host.' The 'security governor' checks the 'host database' and forwards the SC of the 'host' to the 'host registry.' If there is no available SC for the 'host', the 'security governor' contacts the 'host' to establish a trust and issue the SC. Once the SC is issued and passed to the 'host registry', the Service is allocated to the 'host' and a confirmation message forwarded to the 'host'.  
 
\section{Advantage and Disadvantages} \label{opportunities}
In this section we describe benefits and limitation of our MOMCC architecture.
\subsection{Advantages}

$ \bullet $ \textbf{Increased Resource Availability:} By leveraging large number of nearby mobile devices in public places like shopping mall, cinema, and airport the Service availability is increasing noticeably.

$ \bullet $ \textit{Unskilled Hosting:} Mobile owner does not need IT skills to host and run Services, because majority of low-level communication and negotiation can be done automatically without user inference, while the service implementation and maintenance is provided by Service developer. 

$ \bullet $ \textit{Enhanced Security and Reliability:} Nearby mobile devices can directly communicate using WLAN without entering the risky channel of Internet which can often save energy too. The Service provider is unable to identify its consumer(s) which dissuade malicious developers from violating end-users privacy. The Service provider also is discouraged to add malicious code to the Service since the Service governor verifies its credibility. 

$ \bullet $ \textit{Reduced Long WAN Latency:}Exploiting resources of devices in vicinity magnificently reduces the long WAN latency. For computation-intensive applications utilizing Bluetooth or WLAN technologies not only reduced the latency, but also save more energy, while communication-intensive applications are likely using cellular networks \cite{Perrucci}.

$ \bullet $ \textit{Increased Low-cost Resources:} Utilizing computing resources of nearby mobile hosts, does not need an upfront investment since they have already being acquired by their owners. The ownership cost also will be shrunk with resource sharing and earning.

$ \bullet $ \textit{Green Computing:} Contemporary mobile devices have become a luxury device for public and are usually maintain for lightweight tasks such as web browsing, social networking, or basic gaming. Therefore, unused resources can be utilized by heavy applications toward a greener computing.

\subsection{Disadvantages}
$ \bullet $ \textit{Host Computing Limitation:} One of the drawbacks of MOMCC architecture is that complexity of Services is highly dependent on computing capabilities of hosting devices which are not very high. Heavy enterprise applications might found MOMMC difficult to utilize. In our future work, we will utilize resource-rich stationary devices to serve enterprise and resource-intensive applications.

$ \bullet $ \textit{Fine granularity of Services:} Due to limitation of host, Services are often fine in granularity which causes extra execution overhead on mobile hosts. Such overhead prolongs execution time and increases communication traffic as well which will be investigated in our future work.

$ \bullet $ \textit{Unusable in Remote Area:} This architecture is highly dependent on nearby computing devices and will be affected in the remote environment such as mountain or jungles where less mobile devices exist.

$ \bullet $ \textit{No Offline Usability} Applications built on MOMCC architecture, like other Service-based software, are highly dependent on networking and communication with external devices. Application will not be useful in offline mode.

\section{Conclusions}\label{conclusions}
In this paper we employ SOA and propose a Market-Oriented architecture for mobile cloud computing which is the first of its own to the best of our knowledge. This architecture consists of four major entities, namely Service developer, host, governor, and requester. Service developer and host are clearly detached from each other to not only facilitate and encourage Service hosting by unskilled mobile users, but also to increase privacy of the Service requester (consumers). A cloud of mobile devices including smartphones, Tablets, and sundry mobile devices is created where devices with heterogeneous platforms, hardware, and manufacturers can coexist and collaborate. We encourage Service development and hosting by providing monetary incentive for programmers and mobile owners to stimulate mobile Service hosting. Based on Service governor responsibilities we argue that mobile network operators are likely the best candidate to serve as the Service governor, because they are centralized, well-established, and reputed organizations that have been serving mobile users since long and could establish high degree of trust with users. Successful mobile Service hosting architecture can be utilized in various domains such as supply chain management in which various organizations (e.g. billing and transport) can collaborate to perform a business activity. A Service hosted on a driver's mobile can notify customer orders and update delivery scheduling to the recipient. However, MOMCC architecture is more suitable for computing-intensive tasks since different hosts share their computational resources. Data-intensive tasks are less likely addressable in this architecture.

In our future work, we will implement the proposed MOMCC architecture and accommodate data-intensive and complex enterprise applications by employing resourceful computing devices in near and far distance. Utilizing certified surrogate machines (by security governor) and Cloud infrastructures will enable Service developer to build computational- and data-intensive applications.

\section*{Acknowledgment}
This work is funded by the Malaysian Ministry of Higher Education under the University of Malaya High Impact Research Grant  UM.C/HIR/MOHE/FCSIT/03.

\bibliography{mydb}
\bibliographystyle{IEEEtran}

\end{document}